\documentclass[aps,pra,twocolumn,floatfix,showpacs]{revtex4-1}
\usepackage{amssymb,amsmath,graphicx,times}
\def\RTN{\scriptscriptstyle RTN}
\begin{document}
\title{Dynamics of quantum correlations in colored environments}
\author{Claudia~Benedetti }\email{claudia.benedetti@unimi.it}
\affiliation{Dipartimento di Fisica dell'Universit\`{a} degli 
Studi di Milano, I-20133 Milano, Italy}
\author{Fabrizio~Buscemi }\email{fabrizio.buscemi@unimore.it}
\affiliation{ ``E. De Castro'' Advanced Research Center on Electronic 
Systems (ARCES), Universit\`{a} di Bologna, Via Toffano 2/2, I-40125 Bologna,
Italy} 
\author{Paolo~Bordone}\email{paolo.bordone@unimore.it}
\affiliation{Dipartimento di Scienze Fisiche, Informatiche e Matematiche, 
Universit\`{a} di Modena e Reggio Emilia, and Centro S3, CNR-Istituto di 
Nanoscienze, via Campi 213/A,  Modena I-41125, Italy}
\author{Matteo G. A. ~Paris}\email{matteo.paris@fisica.unimi.it}
\affiliation{Dipartimento di Fisica dell'Universit\`{a} degli Studi di
Milano, I-20133 Milano, Italy}
\affiliation{CNISM -- Udr Milano, I- 20133, Milano, Italy}
\date{\today}
\begin{abstract}
We address the dynamics of entanglement and  quantum discord for two non
interacting qubits initially prepared in a maximally entangled state and
then subjected to a classical colored noise, i.e. coupled with an
external environment characterized by a noise spectrum of the form
$1/f^{\alpha}$. More specifically, we address systems where the Gaussian
approximation fails, i.e. the sole knowledge of the spectrum is not enough
to determine the dynamics of quantum correlations. We thus investigate the 
dynamics for two different configurations of the environment: in the first case the noise
spectrum is due to the interaction of each qubit with a single bistable
fluctuator with an undetermined switching rate, whereas in the second
case we consider a collection of classical fluctuators with fixed
switching rates. In both cases we found analytical expressions for 
the time dependence of entanglement and quantum discord, which may
be also extended to a collection of flcutuators with random switching
rates. The environmental noise is introduced by means of
stochastic time-dependent terms in the Hamiltonian and this allows us to 
describe the effects of both separate and common environments. 
We show that the non-Gaussian character of the noise may lead
to significant effects, e.g. environments with the same power
spectrum, but different configurations, give raise to opposite behavior
for the quantum correlations.  In particular, depending on the
characteristics of the environmental noise considered, both entanglement
and discord display either a monotonic decay or the phenomena of sudden
death and revivals. Our results show that the microscopic structure of
environment, besides its noise spectrum, is relevant for the dynamics of
quantum correlations, and may be a valid starting point for the
engineering of non-Gaussian colored environments.
\end{abstract}
\maketitle
\section{Introduction}\label{sez1}
Entanglement and discord describe remarkable features of quantum
systems. Indeed, they are closely related to the amount of quantum
correlations contained in the system, coming either from non-separability
or the impossibility of local discrimination \cite{rev12}.  
Beyond their specific
role in fundamental physics, entanglement and discord has been also
recognized as resources for quantum technology, e.g. for the processing
of quantum information and for the effective implementation of quantum
enhanced protocols. 
In particular, in the last decade it
has been recognized \cite{zurek,maziero,datta,werlang} that separable
mixed states may represent a resource, if they show a nonzero quantum
discord. In fact, mixed separable states with non-zero discord have been
exploited to achieve a speed-up for certain computational tasks compared
to classical states~\cite{datta2,lanyon}.  This is true also for
continuous variable systems, where Gaussian quantum discord has been
introduced \cite{GQD10,Ade10}, measured \cite{expD1}, and exploited for
quantum enhanced protocols \cite{expD2,expD3}.
\par
An essential ingredient to exploit the quantumness of
a physical system is the preservation of its coherent time evolution. On
the other hand, the unavoidable interaction with its environment usually
destroys coherence and quantumness \cite{giulini,Zurek2}, and in turn its
use for quantum technology. For these reasons, much attention has been
devoted to the analysis, characterization and control of the dynamics of
quantum correlations in different physical systems \cite{hu1}, including quantum
optics \cite{cvqc1,cvqc2,cvqc3,cvqc4}, nuclear magnetic
resonance~\cite{Warren,Oliveira},
nanophysics~\cite{buscemitelep,buscemidem} and
biology~\cite{Sarovar,Fassioli}.  
\par
For bipartite open quantum systems interacting with a quantum
environment, entanglement  and discord may exhibit peculiar features
such as sudden death and transitions, revivals, and
trapping~\cite{yu,yu2,yang,cao,s1,ma}.  Such phenomena have been linked
either to direct~\cite{Ficek} or indirect~\cite{Mazz} effective
two-qubit interactions. For non-interacting qubits they are due to the
non-Markovian nature of the environment~\cite{Comp2}, which results in
the transfer of correlations back and forth from the two-qubit system to
the various parts of the total system. In this framework, it was shown \cite{Mazz}
that for a common bosonic environment the entanglement between two qubits is more robust 
against decoherence, and its dynamics is faster, than for independent environments 
with the same spectrum. The effect of the noise spectrum on the dynamics 
of quantum discord has been also analyzed \cite{zhang12}. Recently, revivals of  quantum
correlations have been found also for quantum system coupled to
classical sources~\cite{lofranco,Zhou,pal12} and have been connected to a
quantifier of non-Markovianity for the dynamics of a single-qubit.
Indeed, it was proven that a classical noise can mimic,  without loss of
generality, a quantum environment not affected by the system or
influenced in a way that does not result in back-action~\cite{lofranco}.
\par
Among the class of open quantum systems interacting with a classical 
environment, a particular attention has been devoted to systems made of 
two qubits subjected to a classical source of random telegraph noise
(RTN)~\cite{s2,wold,Zhou,Z2,benedetti2,Lof2}, namely interacting with 
a bistable fluctuator, randomly switching between its two states with 
a given rate $\gamma$. Depending upon the ratio between the switching 
rate and the system-environment coupling, the dynamics of a quantum 
system may exhibit Markovian or non-Markovian behavior. In fact, 
the great interest in RTN is due to the fact that it is able to model
environmental fluctuations  appearing in many nanodevices based on
semiconductors, metals and superconductors~\cite{Duty,eroms,parman91,rogers84,rogers85,peters99}.
Furthermore, it  also represents the basic building block to describe
noises of the type $1/f^{\alpha}$, which are responsible for decoherence
in quantum solid-state devices~\cite{weissman,meno,tsai,paladino,Fac2,Bukard}.
This kind of noise spectra stem from the collection of random 
telegraph source with different switching rates and are usually 
referred to as {\em colored} spectra. The color of the noise depends upon the 
value of the parameter $\alpha$ \cite{nota}. 
For $\alpha=1$ the so-called pink ${1}/{f}$ noise  is found, which is
obtained from a set of random telegraph fluctuators weighted by the
inverse of the switching rate. Another interesting case is the
${1}/{f^2}$ spectrum, also called brown noise from its relation to a
Brownian motion. 
\par
Environments characterized by $1/f^{\alpha}$ noise spectra
usually arise when a system is coupled to a large number of bistable
fluctuators, with a specific distribution of their switching rates.
Upon considering a collection of fluctuators, the colored noise may be
implemented by means of a linear combination of sources of RTN, each
characterized by a specific switching rate chosen from a suitable
distribution. On the other hand, the same
spectrum can be obtained if we consider a single fluctuator with a
random switching rate. As already observed by several authors \cite{paladino,galperin06,bergli}
different microscopic  configurations of the environment leading to
the same spectra, may correspond to different physical phenomena, 
e.g. different evolutions for the quantum correlations.
In these cases, i.e. when the knowledge of the noise spectrum is not
sufficient to describe decoherence phenomena, the noise is referred to be non-Gaussian.
In particular,  we investigate the time evolution of entanglement and
discord of two initially entangled non-interacting qubits coupled to a
classical environment described either by means of a single random
fluctuator or a collection of RTN sources.  For the model of a single
fluctuator with random switching rate, the correlations decay with a
damped oscillating behavior. In the configuration with many bistable
fluctuators, pink noise leads to a monotonic decay of entanglement and
discord, while the presence of brown noise induces phenomena of sudden
death and revivals. We ascribe this discrepancies to the different
number of decoherence channels in the two configurations.  In other
words, the time-evolution of the system is determined not only by the
spectrum of the environment, but also by its configurations, i.e. to its
microscopic structure.
\par  
For both the configurations, the dynamics of the two qubits is ruled by
a stochastic Hamiltonian with time dependent coupling.  The average of
the time-evolved states over the switching parameters describes the
evolution of the two-qubit state under the effect of the noise. 
Upon a suitable choice of stochastic
time-dependent terms in the Hamiltonian, we are able to describe the
effects of both separate and common environments: in the former each
qubit is locally coupled to a random external signal, while in the
latter both qubits are subject to the effect of a common classical
environment.  The interest in systems interacting with a common
environment arises from recent experiments in cavity QED \cite{moe07}
and circuit QED \cite{sillanpa07,majer07} where reservoir-mediated
interaction between two qubits is at the heart of entanglement
generation.
\par
Environmental effects due to a collection of bistable fluctuators have
been experimentally observed in nanoscale electronic devices, where the
single electron tunnelling turns out to be affected by charge
fluctuations \cite{ludvi84,kogan84}.  Our results may be thus used as a
valid guideline to analyze decoherence phenomena in Josephson circuits
subject to noise stemming from fluctuating background charges and flux
\cite{mx1,mx2,mx3,mx4}.  Other systems where the noise is non-Gaussian
and comes from a collection of two-level fluctuators include  silicon
\cite{Joh00} and magnetic ones \cite{Raq00}, as well as vortex matter
\cite{Jun03}. 
\par
The paper is organized as follows: in Section 2, we briefly review the
definition of negativity as a measure of entanglement, and of quantum
discord. In Section 3 we present the physical model for two qubits 
interacting with a classical environment, being represented either 
by a single random bistable fluctuator or a collection of fluctuators.
Section 4 reports results for the different configurations, whereas 
Section 5 closes the paper with a discussion and some concluding remarks.
\section{Entanglement and quantum discord}
In this section we briefly review the concepts of entanglement and
quantum discord between two qubits. In particular we evaluate
entanglement by means of negativity\cite{negat}, which is given by:
\begin{equation}
 N=2\left|\sum_i\lambda_i^{-}\right|\label{negativity}
\end{equation}
where $\lambda^-_i$ are the negative eigenvalues of the partial
transpose of the system density matrix. Note that the negativity is
bound between zero, for separable states, and one, for maximally
entangled states. The concept of negativity has recently been  extended
to the case of tripartite systems of identical particles \cite{neg_trip}.
\par
The quantumness of correlations of bipartite a system may be also 
quantified by discord \cite{zurek,vedral}, namely the difference
between the total and the classical correlations in a system.  The
quantum mutual information quantifies the total correlations in a system
and is defined as:
$ I=S(\rho^A)+S(\rho^B)-S(\rho)$ where $\rho^{A(B)}$ is the partial
trace of the total bipartite system $\rho$ and
$S(\rho)=-\text{Tr}\rho\log_2(\rho)$ is the von Neumann entropy.  The
classical correlations are evaluated by means of the expression
\cite{vedral} $
C=\max_{\{\Pi_j\}}\left[S(\rho^A)-S(\rho^A|\{\Pi_j\})\right]$
where $\{\Pi_j\}$ are projective measurements on subsystem $B$ and
$S(\rho^A|\{\Pi_j\})=\sum_jp_j S\left(\rho^A_j\right)$, with
$\rho^A_j=\text{Tr}_B[\Pi_j\rho\Pi_j]/\text{Tr}[\Pi_j\rho\Pi_j]$ is the
remaining state of $A$ after obtaining outcome $j$ on $B$. Therefore the
quantum discord is the difference between the mutual information and the
classical correlation: 
\begin{align}
 Q=I-C.
\end{align}
Usually, the evaluation of quantum discord is not an easy task, since it
involves an optimization problem. However, for two-qubit systems
described by a density matrix of the form (from now on referred to as
states with a X shape)
$\rho=\frac{1}{4}\left(\mathbb{I}+\sum_{j=1}^3c_j\sigma_j\otimes\sigma_j\right)$,
where $\sigma_j$ are the three Pauli matrices, 
the optimization 
may be carried out analytically \cite{luo}, leading to
\begin{align}
 Q=&\frac{1}{4}[(1-c_1-c_2-c_3)\log_2(1-c_1-c_2-c_3)\nonumber\\
&+(1-c_1+c_2+c_3)\log_2(1-c_1+c_2+c_3)\nonumber\\
&+(1+c_1-c_2+c_3)\log_2(1+c_1-c_2+c_3)\nonumber\\
&+(1+c_1+c_2-c_3)\log_2(1+c_1+c_2-c_3)]\nonumber\\
&-\frac{1-c}{2}\log_2(1-c)-\frac{1+c}{2}\log_2(1+c)\label{discordluo},
\end{align}
where $c:=\max\{|c_1|,|c_2|,|c_3|\}$.
\section{The physical model}\label{model}
We address quantum correlations, both entanglement
and discord, between two non-interacting qubits subjected to noisy
environments. Our analysis concerns environments with noise 
spectra of the form $1/f^\alpha$, which are realized by different 
configurations of bistable fluctuators.
The two qubits are initially
prepared in the Bell state
$|\phi^+\rangle=(|00\rangle+|11\rangle)/\sqrt{2}$. The interaction
between the qubits and the environment can be either local or global,
i.e. we examine both the case of independent environments acting locally
on each qubit and of a common environment affecting the two qubits. If
we set $\hbar=1$ and adopt the spin notation, the two-qubit Hamiltonian
describing the interaction with a single fluctuator is given by
\begin{equation}
 H(t)=H_A(t)\otimes\mathbb{I}_B+\mathbb{I}_A\otimes H_B(t)\label{H2qubit},
\end{equation}
where $H_{A(B)}$ is the Hamiltonian of a single qubit subject to a 
classical time-dependent noise which affects the transition amplitude parameter $c_{A(B)}(t)$:
\begin{equation}
 H_{A(B)}(t)=\epsilon \mathbb{I}_{A(B)}+\nu c_{A(B)}(t)\sigma_{x\;A(B)},\label{h1qubit}
\end{equation}
with $\epsilon$ the
qubit energy in absence of noise (energy degeneracy is assumed), $\mathbb{I}_{A(B)}$  the identity
matrix for subspace $A(B)$, $\nu$ is the coupling constant between the
system and the environment, $\sigma_x$ the Pauli matrix.  If the
time-dependent coefficient $c_{A(B)}(t)$ can randomly flip between two
values $c(t)=\pm1$ with a fixed rate $\gamma$, then Eq. \eqref{h1qubit}
describes a qubit subject to a random telegraph noise
\cite{s2,bergli,Zhou,benedetti2,Lof2,benedetti3}. This Hamiltonian,
extended to describe qudits, has also been used to analyze the time
evolution of entanglement of a continuous-time quantum walk of two
indistinguishable particles on a one-dimensional lattice
\cite{benedetti1}.  The Hamiltonian \eqref{H2qubit} is stochastic due to
the random nature of the noise parameter $c(t)$. For a specific choice
of $c(t)$, the total system evolves  according to the evolution operator
$e^{-i\int H(t')\:dt'}$, with positivity ensured by the very structure of 
the Hamiltonian in Eq. (\ref{h1qubit}) \cite{daf04}. 
By averaging the global state over different
realizations of the sequences of  $c(t)$, the two-qubit mixed state is
obtained. 
\par
In order to reproduce the $1/f^{\alpha}$ spectrum, the
single RTN frequency power density must be integrated over the switching
rates $\gamma$ with a proper distribution:
\begin{align}
 S_{1/f^{\alpha}}(f)=\int_{\gamma_1}^{\gamma_2}S_{\RTN}(f,\gamma)p_{\alpha}(\gamma)\,d\gamma\label{spectrum},
\end{align}
where $S_{\RTN}(f,\gamma)$ is the random telegraph noise frequency
spectral density with lorentzian form $S_{\RTN}(f,\gamma)=4\gamma/(4\pi^2
f^2+\gamma^2)$.  The integration is performed between a minimum and a
maximum value of the switching rates, respectively $\gamma_1$ and
$\gamma_2$.  $p(\gamma)$ is the switching rate distribution and takes a
different form depending on the kind of noise:
\begin{align}\label{distrib}
p_{\alpha}(\gamma)=\left\{\begin{array}{ll}
 \frac{1}{\gamma\,\ln(\gamma_2/\gamma_1)}&\alpha=1\\
 & \\
\frac{(\alpha-1)}{\gamma^{\alpha}}
 \left[\frac{(\gamma_1\gamma_2)^{\alpha-1}}
 {\gamma_2^{\alpha-1}-\gamma_1^{\alpha-1}}\right]&1<\alpha\leqslant2 \,.
 \end{array}
\right.
\end{align}
It follows that, in order to simulate a frequency spectrum proportional
to $1/f^{\alpha}$, the switching rates must be selected from a
distribution proportional to $1/\gamma^{\alpha}$.  When  the integration in Eq. 
\eqref{spectrum} is performed, the spectrum has the requested
$1/f^{\alpha}$ behavior in a frequency interval, so that every frequency belonging to such interval satisfies the condition
 $\gamma_1\ll f\ll\gamma_2$.
Eq. \eqref{spectrum}
can be obtained either considering a single bistable fluctuator whose
switching rate is randomly chosen from a distribution
$p_{\alpha}(\gamma)$ or from a collection of sources of RTN each with a
switching rate taken from the same $\gamma$-distribution. Even if the
spectrum is the same, the physical systems described are indeed very
different.  
\subsection{$1/f^{\alpha}$ noise from a single fluctuator}
In the case of a single fluctuator, the noise parameter $c(t)$ can only
flip between two values $\pm1$. The difference with the RTN case, is
that here the switching rate is not known a-priori. This means that the
bistable fluctuator is described by a statistical mixture whose elements
are chosen from the ensemble $\{\gamma,p_{\alpha}(\gamma)\}$. This model
describes the physical situation in which each of the two qubits
interacts with a single fluctuator, for separate baths, or with the same 
fluctuator in the case of a common bath.
The qubits are affected only by one source of noise, and therefore only
one decoherence channel is present.\par 
The global system evolves
according to the Hamiltonian \eqref{H2qubit}, with a specific choice of
both the parameter $c(t)$ and of its switching rate. In particular, for
each value of the switching rate picked from a distribution
$p_{\alpha}(\gamma)$, we create a large number of sequences $c(t)$. This
means that we are actually introducing some uncertainty on the rate
$\gamma$. 
This uncertainty gives rise to a $1/f^{\alpha}$ spectrum. For each
selected switching rate, the evolution corresponds to the one of a
bistable fluctuator with a characteristic RTN phase shown to be:
\begin{equation}
 \varphi_{A(B)}(t)=-\nu\int_0^tc(t')dt'\label{phase}
\end{equation}
and characterized by a distribution \cite{paladino,bergli}:
\begin{align}\label{distribution_phi}
 &p(\varphi,t)= \frac{1}{2}e^{-\gamma t}\Bigg\{\nonumber \\ 
 &\times[\delta(\varphi+\nu
 t)+\delta(\varphi-\nu t)] +
 \frac{\gamma}{\nu}[\Theta(\varphi+\nu t)+\Theta(\varphi-\nu
 t)]\nonumber\\
&\times\left[\frac{I_1\left(\gamma t\sqrt{1-(\varphi/\nu t)^2}\right)}{\sqrt{1-(\varphi/\nu t)^2}}
+I_0\left(\gamma t\sqrt{1-(\varphi/\nu t)^2}\right)\right]\Bigg\}
\end{align} 
where $I_v(x)$ is the modified Bessel function and 
$\Theta(x)$ is the Heaviside step function.
For a given $\gamma$ the global system is described by a X-shaped
density matrix obtained by averaging over the noise phase the density
matrix $\rho(\varphi,\gamma,t)$ corresponding to a specific choice of
the parameter $c(t)$ \cite{benedetti3} : 
\begin{align}
\rho(\gamma,t)&=\int \rho(\varphi,\gamma,t) p(\varphi,t)d\varphi\nonumber\\
&=\frac{1}{2}\Big[(1+\beta_{de(ce)})|\phi^+\rangle\langle\phi^+|
\notag \\ &+(1-\beta_{de(ce)})|\psi^+\rangle\langle\psi^+|\Big]\label{rtn}
\end{align}
where  $|\psi^+\rangle=(|01\rangle+|10\rangle)/\sqrt{2}$ is a Bell state
and $\beta_{de}=D^2_{2\nu}(t)$ and $\beta_{ce}=D_{4\nu}(t)$ are
time-dependent coefficients. The function $D_{m\nu}(t)$ represents the
average of the RTN phase factor i.e. 
\begin{equation}
\langle e^{im\varphi(t)}\rangle=\int
e^{im\varphi(t)}p(\varphi,t)\,d\varphi=D_{m\nu}(t),
\end{equation}
where
\begin{equation}
\label{ddtdif}
D_{m\nu}(t)=\left\{ 
\begin{array}{l}e^{-\gamma t}\left[\cosh {\left(\kappa_{m\nu}t\right)} 
+\frac{\gamma}{\kappa_{m\nu}}\sinh {\left(\kappa_{m\nu}t\right)}\right]
 \\ 
 e^{-\gamma t}\left[\cos {\left(\kappa_{m\nu}t\right)} +
 \frac{\gamma}{\kappa_{m\nu}}\sin {\left(\kappa_{m\nu}t\right)}\right]
\end{array}\right . ,
  \end{equation}
for $\gamma \geq m\nu$ and $\gamma \leq m\nu$ respectively, 
where $\kappa_{m\nu}$=$\sqrt{|\gamma^2 -(m\nu)^2|}$ with $m\in\{2,4\}$.
The two-qubit density matrix is obtained by averaging the density 
operator in Eq. \eqref{rtn} over  the switching rates:
\begin{align}
& \overline{\rho}_{de}(t)=
\int_{\gamma_{1}}^{\gamma_{2}}\!\!\int_{\gamma_{1}}^{\gamma_{2}}\!\!\rho_{de}(\gamma,t)\; p_{\alpha}(\gamma_A)p_{\alpha}(\gamma_B)\,\text{d}\gamma_A\text{d}\gamma_B\\
& \overline{\rho}_{ce}(t)=
\int_{\gamma_{1}}^{\gamma_{2}}\!\!\rho_{ce}(\gamma,t)\; p_{\alpha}(\gamma)\,\text{d}\gamma.
\end{align}
Once the average over $\gamma$ is performed, the time-evolved density matrix reads:
\begin{align}
\overline{\rho}_{de(ce)}(t)=
\frac{1}{2}\Big[&(1+\Lambda_{de(ce)})|\phi^+\rangle\langle\phi^+| \notag
\\&+(1-\Lambda_{de(ce)})|\psi^+\rangle\langle\psi^+|\Big]
\label{rho_sing}
\end{align}
where the time-dependent coefficient $\Lambda_{de(ce)}$ can be written as:
\begin{align}
\Lambda_{de}&=\left[\int_{\gamma_1}^{\gamma_2}\, D_{2\nu }\, p_{\alpha}(\gamma)\,
d\gamma\right]^2 \notag\\ 
\Lambda_{ce}&=\int_{\gamma_1}^{\gamma_2}\, D_{4\nu
}\, p_{\alpha}(\gamma)\,d\gamma \label{Lambdas}\end{align}
This means that the quantum system is again described by a
density matrix with a X form.
\subsection{$1/f^{\alpha}$ noise from a collection of fluctuators}
The $1/f^{\alpha}$ noise spectrum can arise from the coupling of a system with
a large number of fluctuators, each characterized by a specific
switching rate, picked from the distribution $p_{\alpha}(\gamma)$
\cite{kuopo,Fac2} in a range $[\gamma_1,\gamma_2]$.  In this case the
random parameters in Eq.\eqref{h1qubit} describes a linear combination
of bistable fluctuators $c(t)=\sum_{j=1}^{N_f}c_j(t)$, where $N_f$ is
the number of fluctuators and we drop the subscript $A(B)$ to simplify
the notation.  Each $c_j(t)$ has a lorentzian power spectrum whose sum
gives the power spectrum of the noise:
\begin{align}
S(f)=\sum _{j=1}^{N_f}S_j(f;\gamma_j)=\sum_{j=1}^{N_f}\frac{\gamma_j}{\gamma_j^2+4\pi^2f}\propto \frac{1}{f^{\alpha}}.
\end{align}
In fact, the sum $\sum _{j=1}^{N_f}S_j(f;\gamma_j)$, with the $\gamma_j$
belonging to the distribution $p_{\alpha}(\gamma_j)$, can be viewed as the
Monte-Carlo sampling of the  integral $N_f\int_{\gamma_1}^{\gamma_2}
S(f)p_{\alpha}(\gamma)\,d\gamma$, which is, but for a constant, the one
in Eq. \eqref{spectrum}. Therefore, in order to obtain a $1/f^{\alpha}$
spectrum, it is necessary that a sufficiently large number of
fluctuators  is considered,  and that the selected $\gamma_j$ are a
representative sample of the distribution $p_{\alpha}(\gamma_j)$ in the
range $[\gamma_1,\gamma_2]$. This means that the minimum number of
fluctuators we can sum-up depends on the range of integration. In fact,
a big number of fluctuators is required to sample the distribution of
the switching rates over a large range, while few fluctuators are
sufficient in the case of a narrow range. Note that we assume that all
the fluctuators have the same coupling constant with the environment,
that is $\nu_j=\nu$ for $j=1\dots N_f$. \par 
The global evolution operator $U(t)\propto e^{-i\sum_j\varphi_j(t)}$, for fixed
values of the parameters associated to each fluctuator, permits to
compute the density matrix of the global system as a function of a total
 noise phase $\varphi(t)=\sum_j\varphi_j(t)$. 
Following the approach used in \cite{benedetti3} to evaluate the dynamics of two qubits subject to a single RTN, the time-evolved density matrix of the system at time $t$ can be expressed as: 
\begin{equation}
 \overline{\rho(t)}=\int \rho(\varphi,t) p_T(\varphi,t)d\varphi, \label{rhomedio1}
\end{equation}
where $p_T(\varphi,t)=\prod_j p(\varphi_j,t)$ is the global
noise phase distribution. The density matrix in Eq. (\ref{rhomedio1}) depends
on the average of the phase factor $e^{im\varphi(t)}$, which can be
computed in terms of the RTN coefficient $D(t)$ of Eq.\eqref{ddtdif}. Note
that for the sake of simplicity, we omitted the subscripts $m\nu$ in writing the $D(t)$ coefficient.\\ We have
\begin{equation}
\langle e^{im\varphi(t)}\rangle=\langle \prod_j e^{im\varphi_j}\rangle=\prod_j D_j(t),\label{prod}
\end{equation}
where the last equality holds since the RTN phase coefficients are independent.
By inserting Eq.\eqref{prod} in Eq.\eqref{rhomedio1}, we evaluate the two-qubit density matrix:
\begin{align}\label{rho_many_fl}
\overline{\rho}_{de(ce)}(t, \, \{\gamma_j\})=
\frac{1}{2}&\Big[(1+\Gamma_{de(ce)})|\phi^+\rangle\langle\phi^+|
\notag \\ &+(1-\Gamma_{de(ce)})|\psi^+\rangle\langle\psi^+|\Big].
\end{align}
The coefficients appearing in the
density matrix are 
\begin{align}\label{Gammas}
\Gamma_{de}&=\prod_j D_{jA}D_{jB}(t)\quad \hbox{with}\quad D(t)=D_{2\nu}(t)
\notag \\
\Gamma_{ce}&=\prod_j D_j(t)\quad  \hbox{with}\quad D(t)=D_{4\nu}(t)\,.
\end{align}
Note that the fluctuators have fixed switching rates $ \{\gamma_j\}$
, $j=1\dots N_f$.\\
\subsection{$1/f^{\alpha}$ noise from a collection of fluctuators with 
random switching rates}
For the sake of completeness, here we  report also
a third scenario which describes an environment with a  $1/f^{\alpha}$
spectrum. This is the case of two qubits subject to a collection of
fluctuators whose switching rates are known with an uncertainty
determined by the $1/\gamma^{\alpha}$ distribution in the finite range
$[\gamma_1,\gamma_2]$. The sources of RTN do not have a fixed $\gamma$,
so they are described by an ensemble of many fluctuators. The dynamics
is evaluated by averaging the two-qubit density matrix for specific
values of the switching rates \eqref{rho_many_fl} over the $\gamma$ in a
range $[\gamma_1,\gamma_2]$: 
\begin{equation}
\overline{\rho}_{de(ce)}(t)=\int_{\gamma_1}^{\gamma_2}
\overline{\rho}_{de(ce)}\left(t, \, \{\gamma_j\}\right)\,p_{\alpha}(\{\gamma_j\})\,d\{\gamma_j\},
\end{equation}
where $p_{\alpha}(\{\gamma_j\})=\prod_jp_{\alpha}(\gamma_j)$ and
$d\{\gamma_j\}=\prod_j d\gamma_j$.
The time evolution also in this case preserves the X shape of 
the density matrix and the evolved state has the same 
functional expression of Eq. \eqref{rho_many_fl}, but with coefficients \begin{equation}
 \Gamma_{de(ce)}\rightarrow\Gamma'_{de(ce)}=\left[\Lambda_{de(ce)}\right]^{N_f}\label{newcoeff}
\end{equation}
where $\Lambda_{de(ce)}$ are  given in Eq.
\eqref{Lambdas}.
\section{Results}
In this section we present analytical expressions for the negativity 
and the quantum discord for qubitq interacting with a single random
fluctuator and a collection of $N_f$ bistable fluctuators, with either
fixed or random switching rates.
\par
As we will see in details in the following, negativity and
discord show the same qualitative behavior in all the analyzed
regimes. The reason for this behavior stays in two features
of the physical systems we are addressing: a) the choice
of a maximally entangled state as the initial state of the 
two qubits; b) the sheer dephasing nature of the interaction.
Overall, these two facts imply that the evolved state is a mixture 
of Bell states. For this class of states discord is a function of 
negativity only, as it follows straightforwardly from Eqs.
(\ref{negativity}) and (\ref{discordluo}).
\subsection{Single fluctuator}
\begin{figure}[htpb]
\includegraphics*[width=0.48\columnwidth]{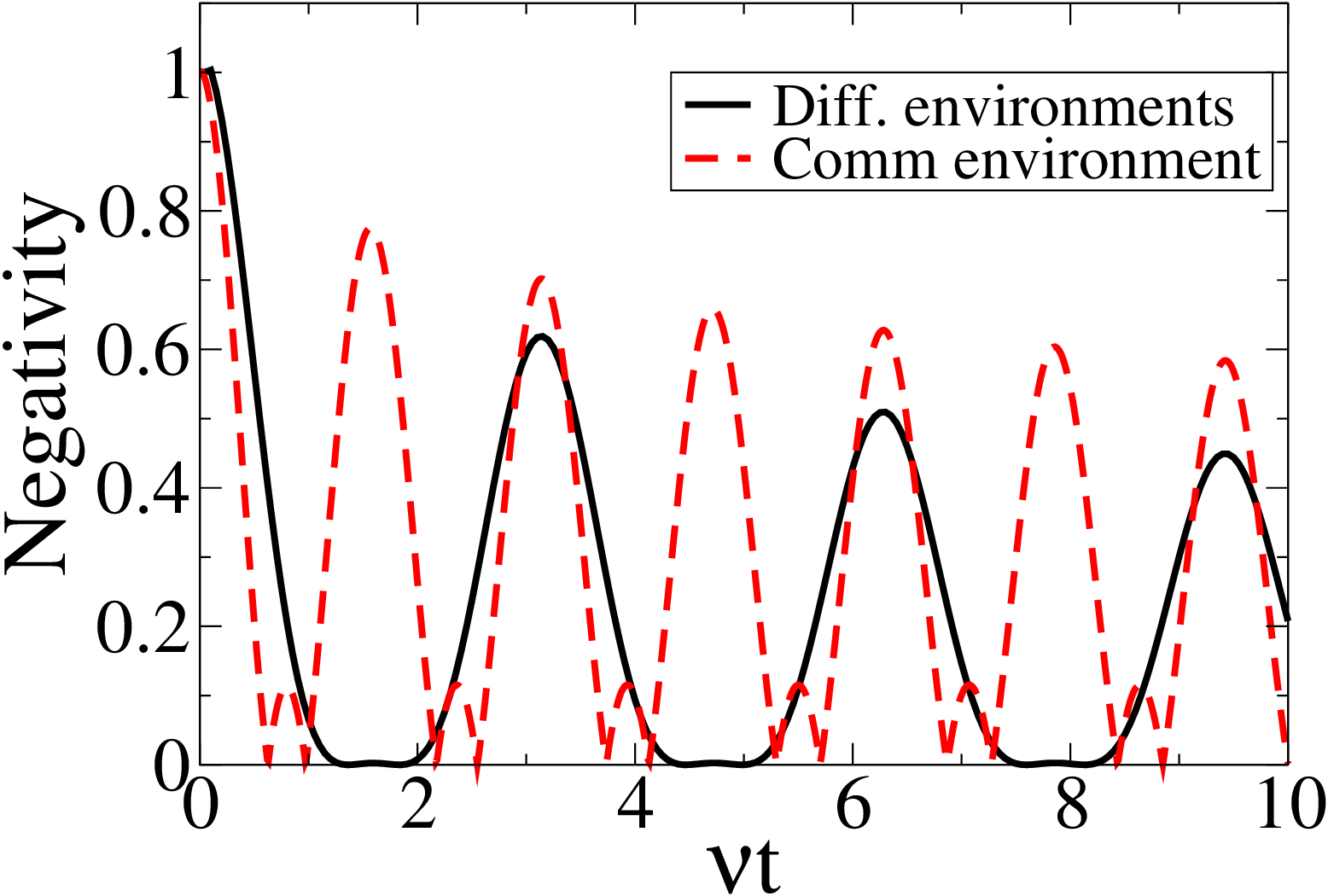}
\includegraphics*[width=0.48\columnwidth]{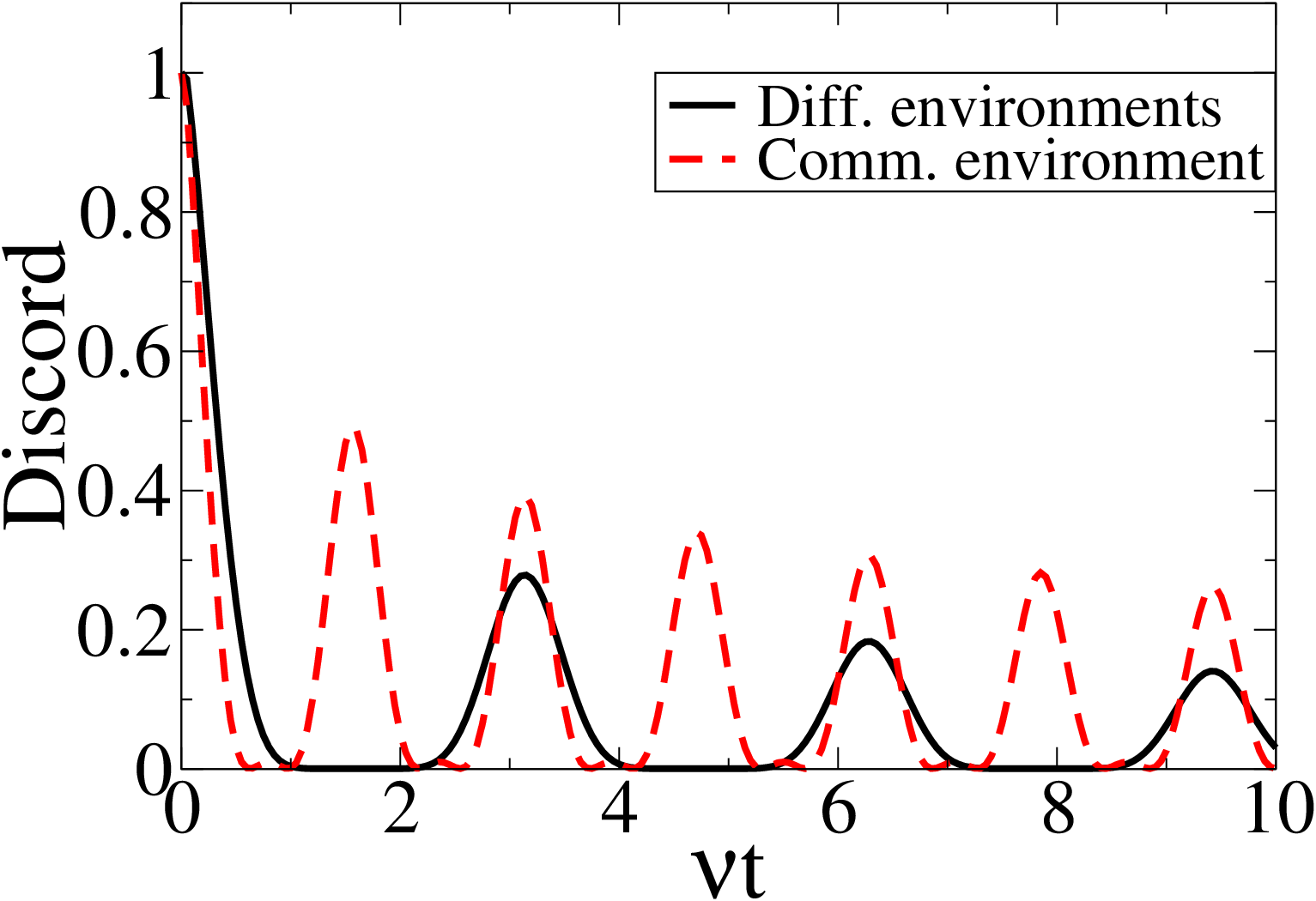}\\ 
\includegraphics*[width=0.48\columnwidth]{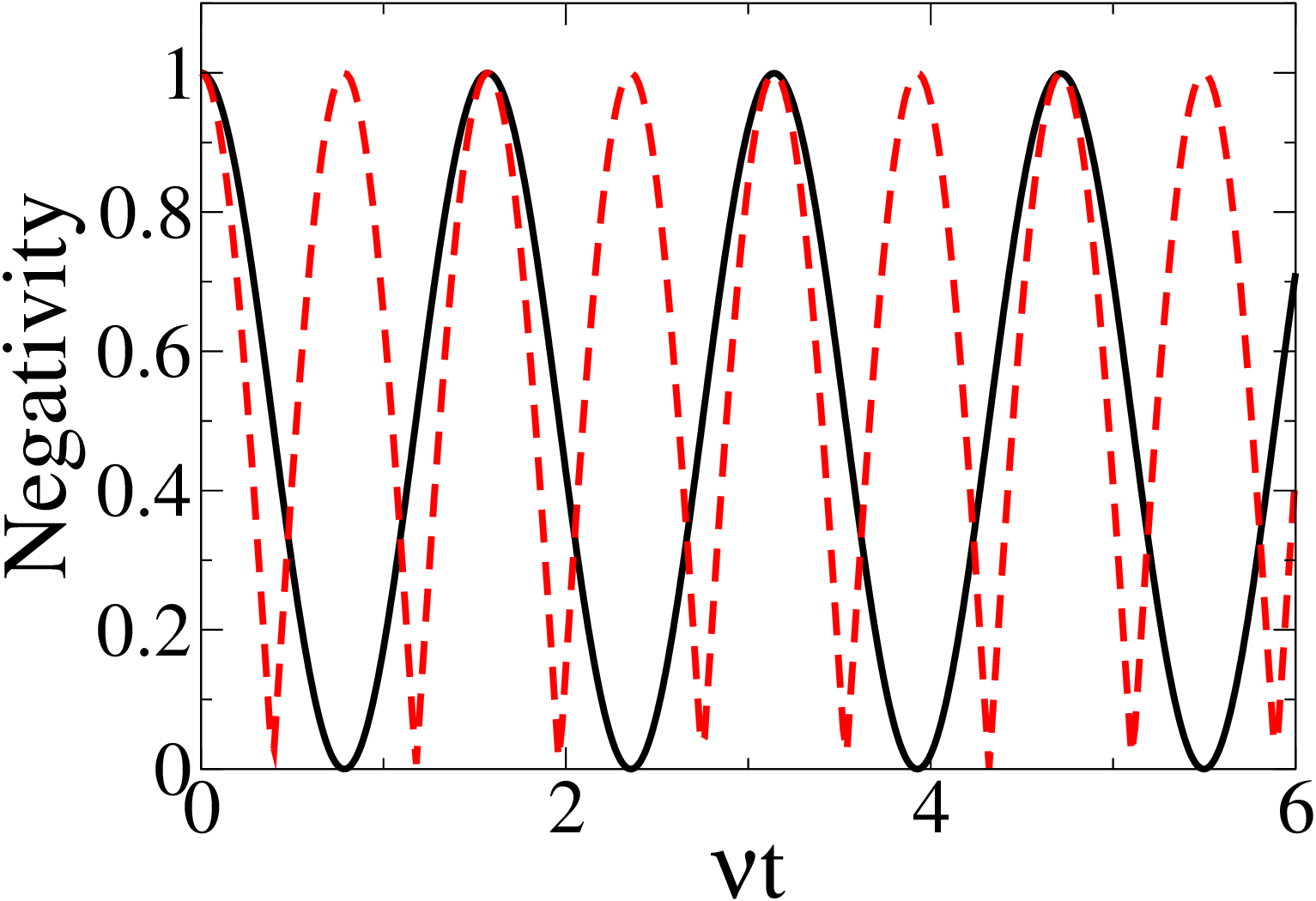}
\includegraphics*[width=0.48\columnwidth]{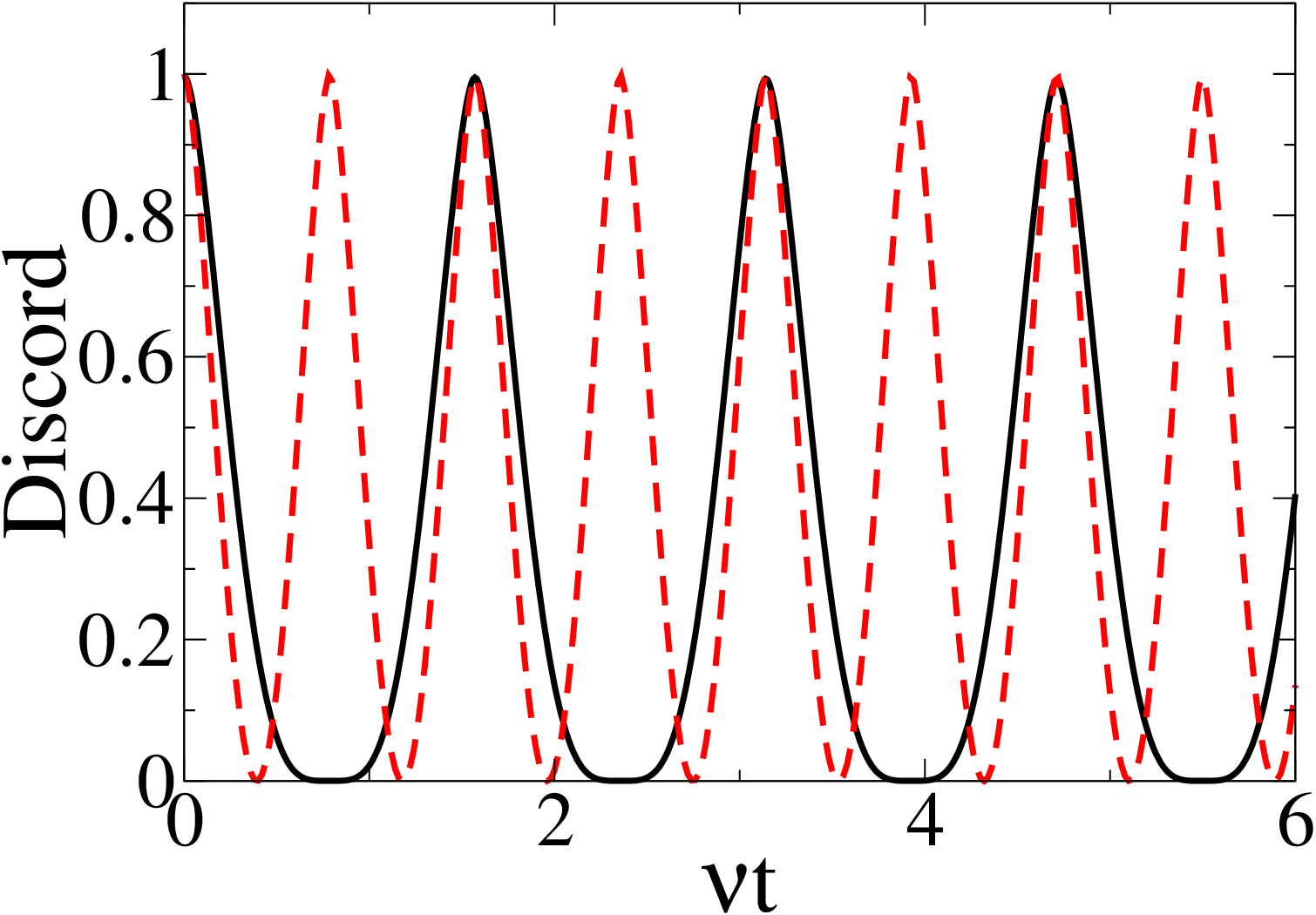}
\caption{ \label{fig2} (Color online) 
Upper panels: Time evolution  of  negativity
(left) and discord (right) for two qubits interacting with a single 
random bistable fluctuator with spectrum $1/f$ for different (solid line) and
common (dashed line) environments when
$[\gamma_1,\gamma_2]/\nu=[10^{-4},10^{4}]$. Bottom panels: Time
evolution  of  negativity (left) and discord (right) for two qubits 
interacting with a single random fluctuator with spectrum $1/f^2$ for different
(solid line)  and common (dashed line) environments when
$[\gamma_1,\gamma_2]/\nu=[10^{-4},10^{4}]$. } 
\end{figure}

In the case of two qubits interacting with a single random fluctuator,
the dynamics depends upon the selected range $[\gamma_1,\gamma_2]$ of
the switching rate.  Following the definition of negativity in Eq.
\eqref{negativity}, the negativity reads:
\begin{align}
N_{de}(t)&=\Lambda_{de}(t)
\\
N_{ce}(t)&=|\Lambda_{ce}(t)|\,,
\end{align}
where the $\Lambda$s are given in Eq. (\ref{Lambdas}). 
Since the density matrix preserves its X shape during 
time evolution, the discord can be evaluated using 
Eq. \eqref{discordluo}:
\begin{equation} 
Q_{de(ce)}(t)= h(\Lambda_{de(ce)})\,,
\label{discsingolo}
\end{equation}
where 
$$ h(x) = \frac{1}{2}\Big[
(1+x)\log_2{\left(1+x\right)} + 
(1-x)\log_2{\left(1-x\right)}
\Big]\,.
$$
The dynamics of $N$ and $D$ is shown in Fig \ref{fig2}, for the cases of
pink and brown noise.  The range of integration is  $[10^{-4},10^4]/\nu$.
Negativity and discord have been obtained analytically, except for the 
integral $\Lambda_{de(ce)}$, which must be computed numerically.
Quantum correlations decay with damped oscillations. In the case of a
qubit subject to different environments with $1/f$ spectrum, the
oscillations have a periodicity of $\pi$.  This periodicity can be
explained by analyzing the analytical expressions of the quantum
correlations. In particular, we note that in the integral of
Eq.\eqref{Lambdas}, the $D_{2\nu}$ functions exhibit damped oscillations
for $\gamma<2\nu$ with periodicity $2\pi/\kappa_{m\nu}$, and for
$\gamma>2\nu$ monotonically decay.  Their weighted superposition leads
to an interference effect that can be summarized as follows: the
oscillating components result in the formation of alternatively positive
and negative peaks spaced by $\pi/2$. On the other hand, the monotonic
decaying components combine to cancel the negative peaks and to preserve
the positive ones. Finally we are left with an oscillating function with
periodicity of $\pi$. The same concept applies in the case of a common
environment, but now the  $D_{4\nu}$ sum up in an oscillating function
with periodicity of $\pi/2$.  \par The $1/f^2$ noise spectrum leads to
oscillating functions of time with periodicity $\pi/2$ and $\pi/4$  for
different and common environments respectively. Again this periodicity
is related to the fact that with such a distribution, the selected
values of $\gamma$ accumulate near the lower value of the frequency
range, thus leading to a beat phenomenon with constructive interference
with the above mentioned periodicity.  If different ranges of
integration are considered, different time-behavior for the quantum
correlations can arise, but we will not analyze this effects in this
paper.  
\subsection{Collection of  fluctuators} 
\begin{figure}[htpb]
\includegraphics*[width=0.48\columnwidth]{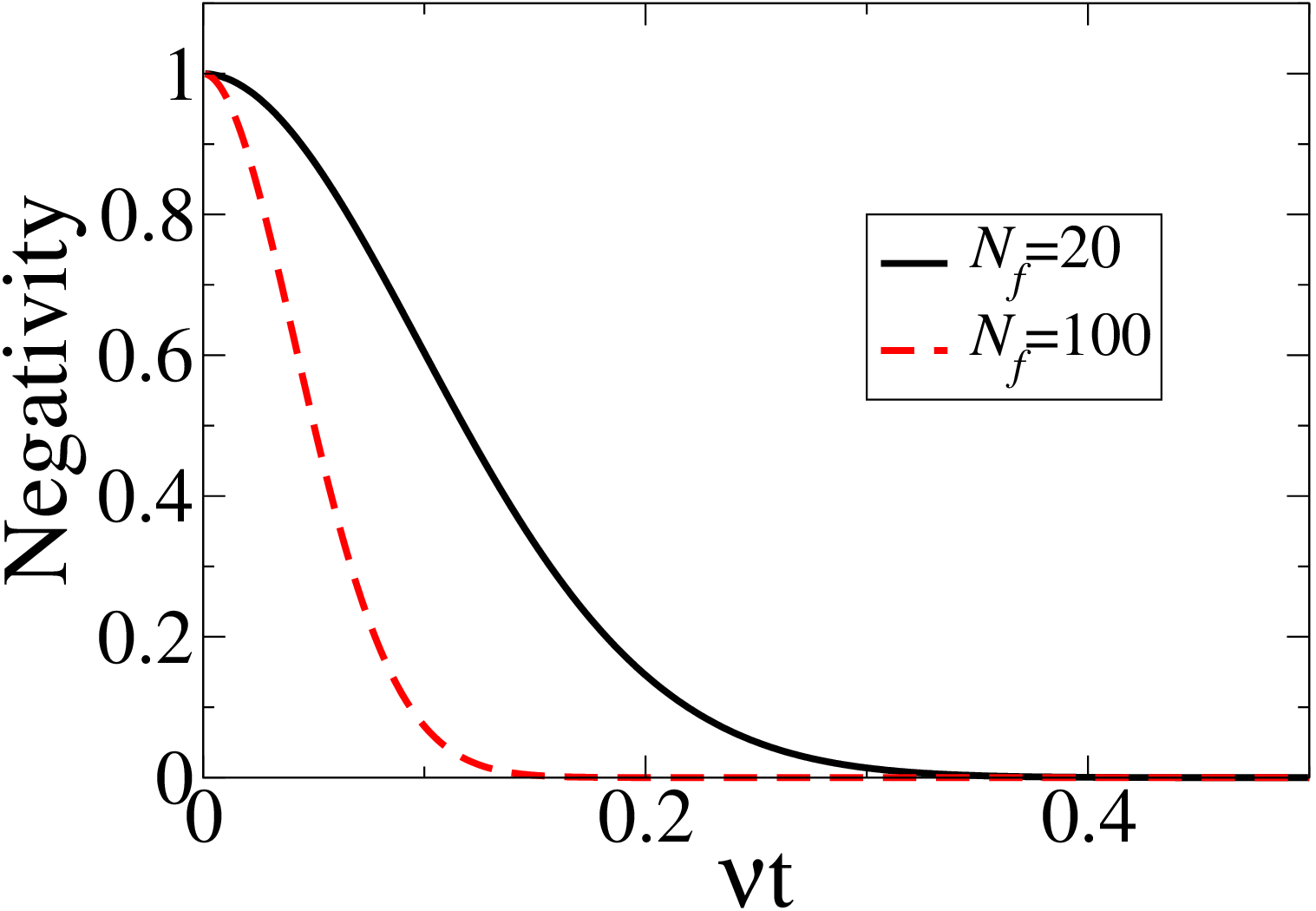} 
\includegraphics*[width=0.48\columnwidth]{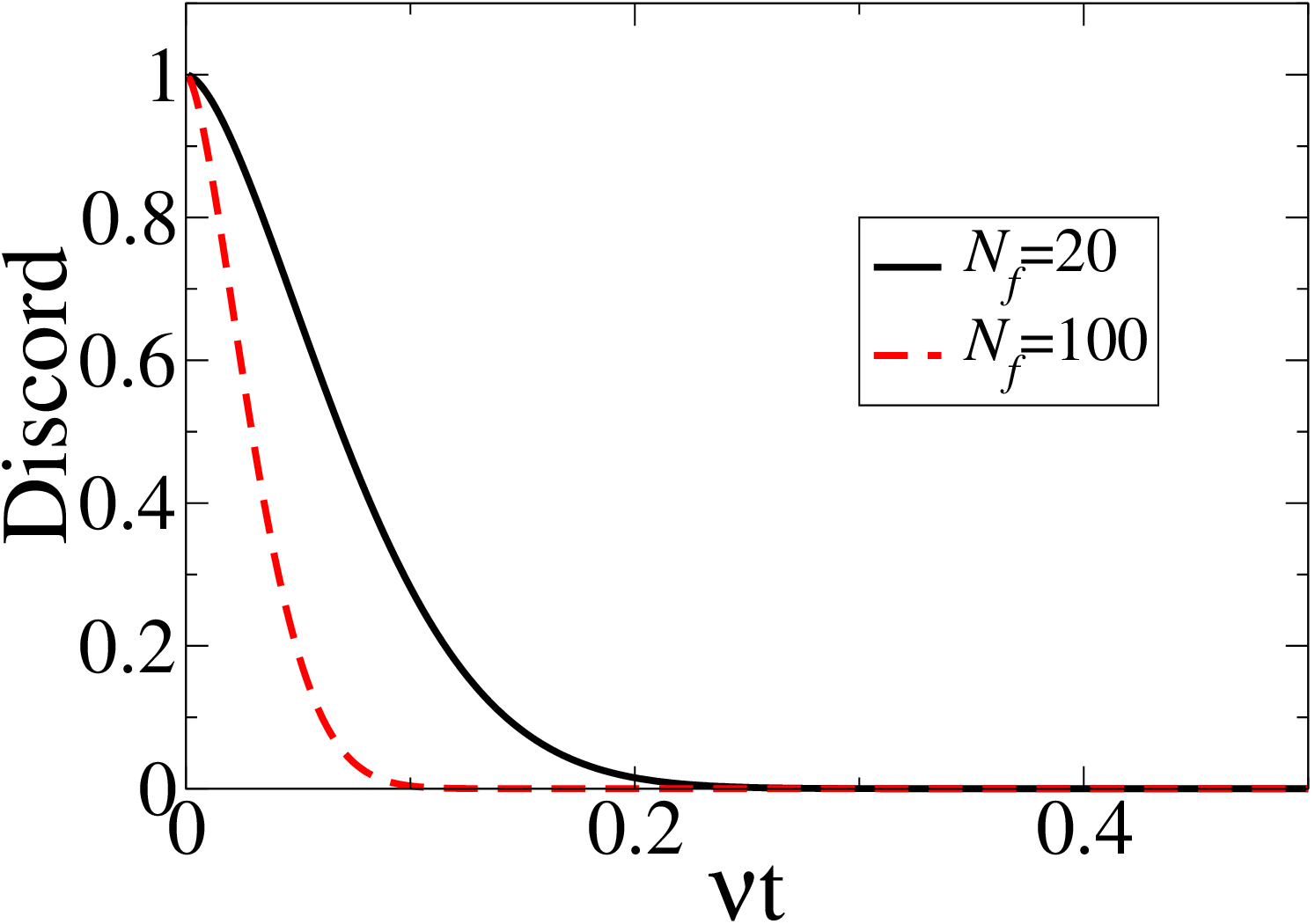}\\
\includegraphics*[width=0.48\columnwidth]{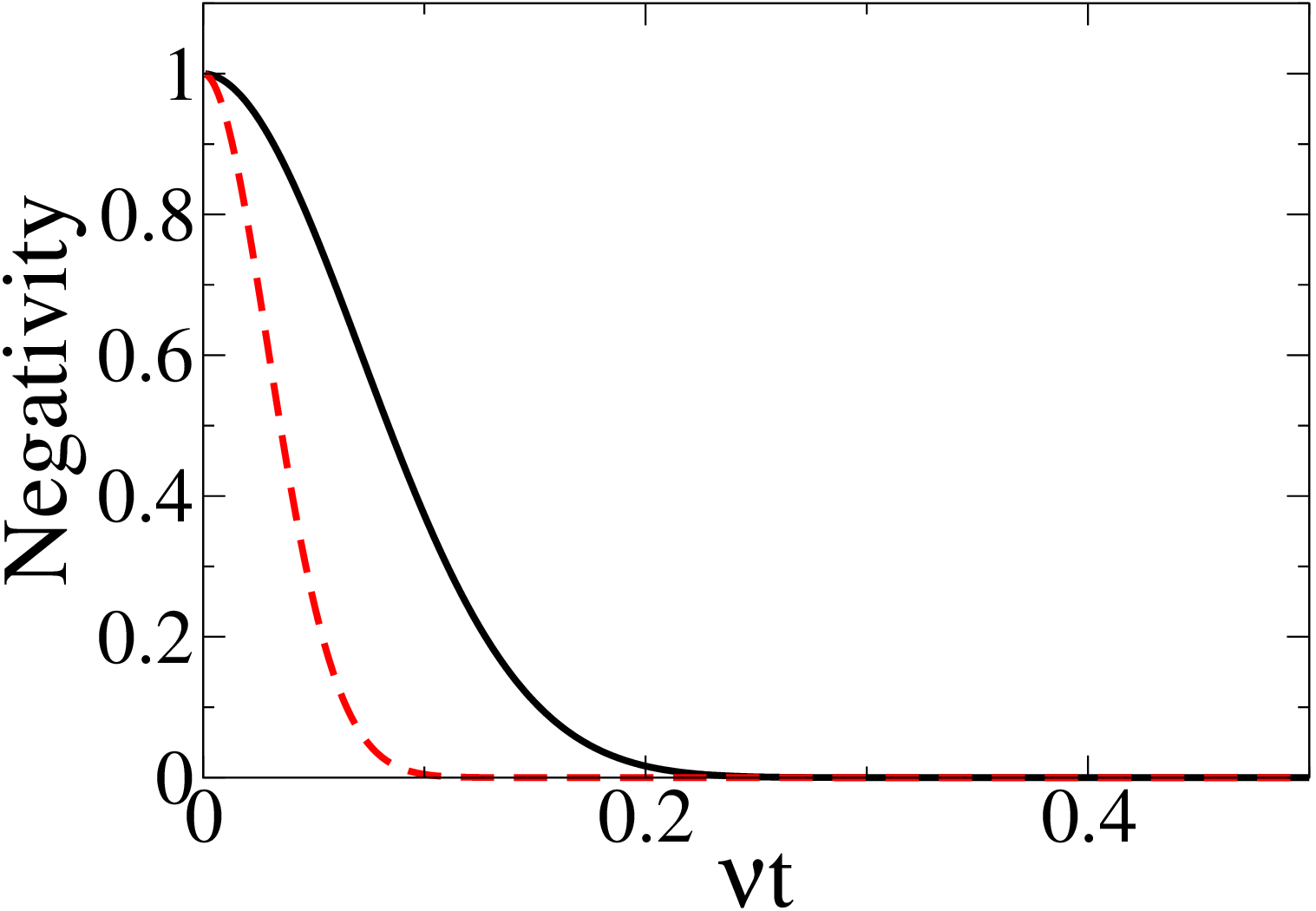}
\includegraphics*[width=0.48\columnwidth]{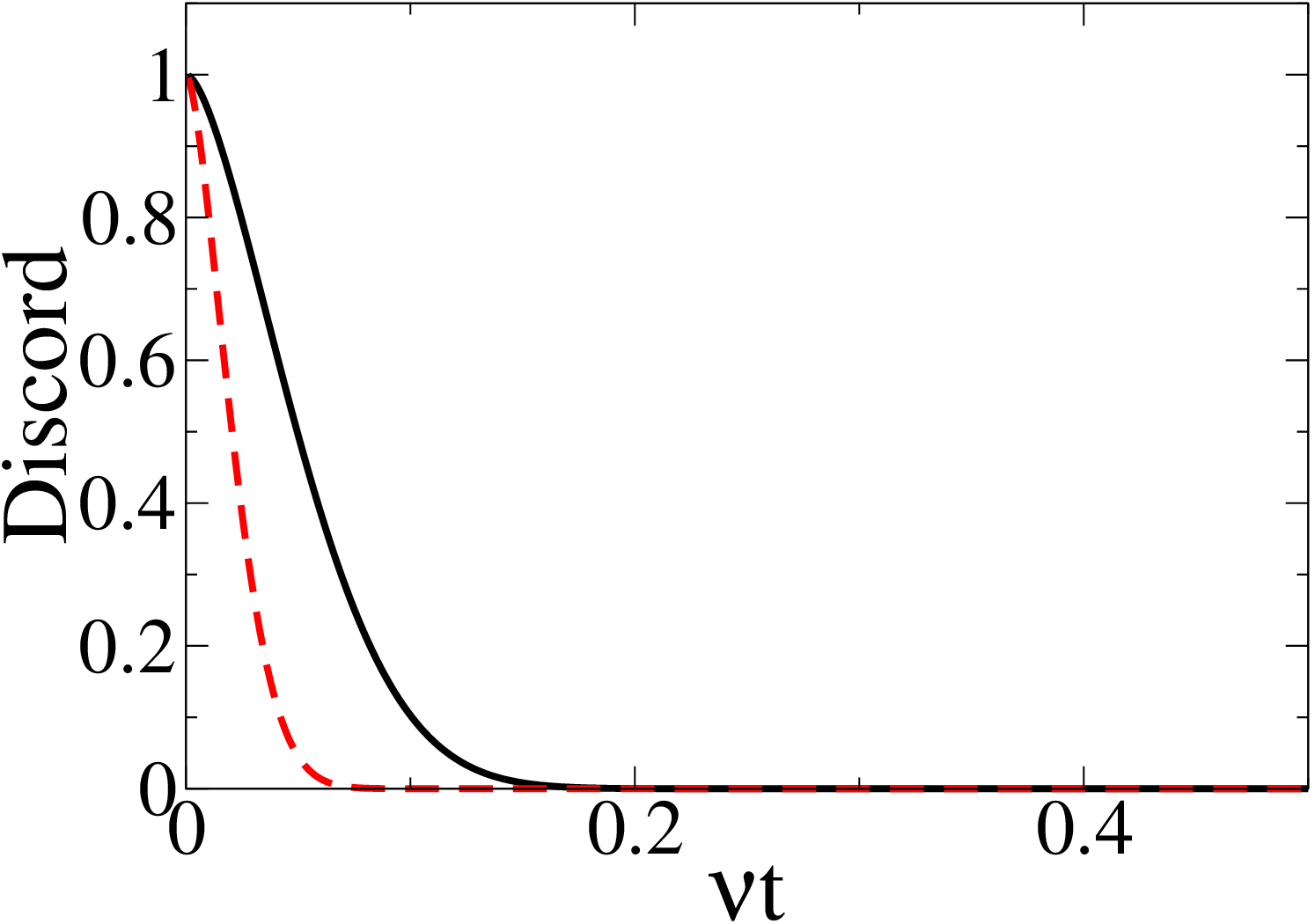}
\caption{\label{fig3} (Color online)
Upper panels: Time evolution  of  negativity (left) and discord (right) 
for two qubits interacting with two independent environments, consisting
in a collection of $N_f$ bistable fluctuators with spectrum $1/f^2$
when $[\gamma_1,\gamma_2]/\nu=[10^{-4},10^{4}]$. Lower 
panels: same as before, but with qubits subject to a  common
environment.}
\end{figure}

\begin{figure}[htpb]
\includegraphics*[width=0.48\columnwidth]{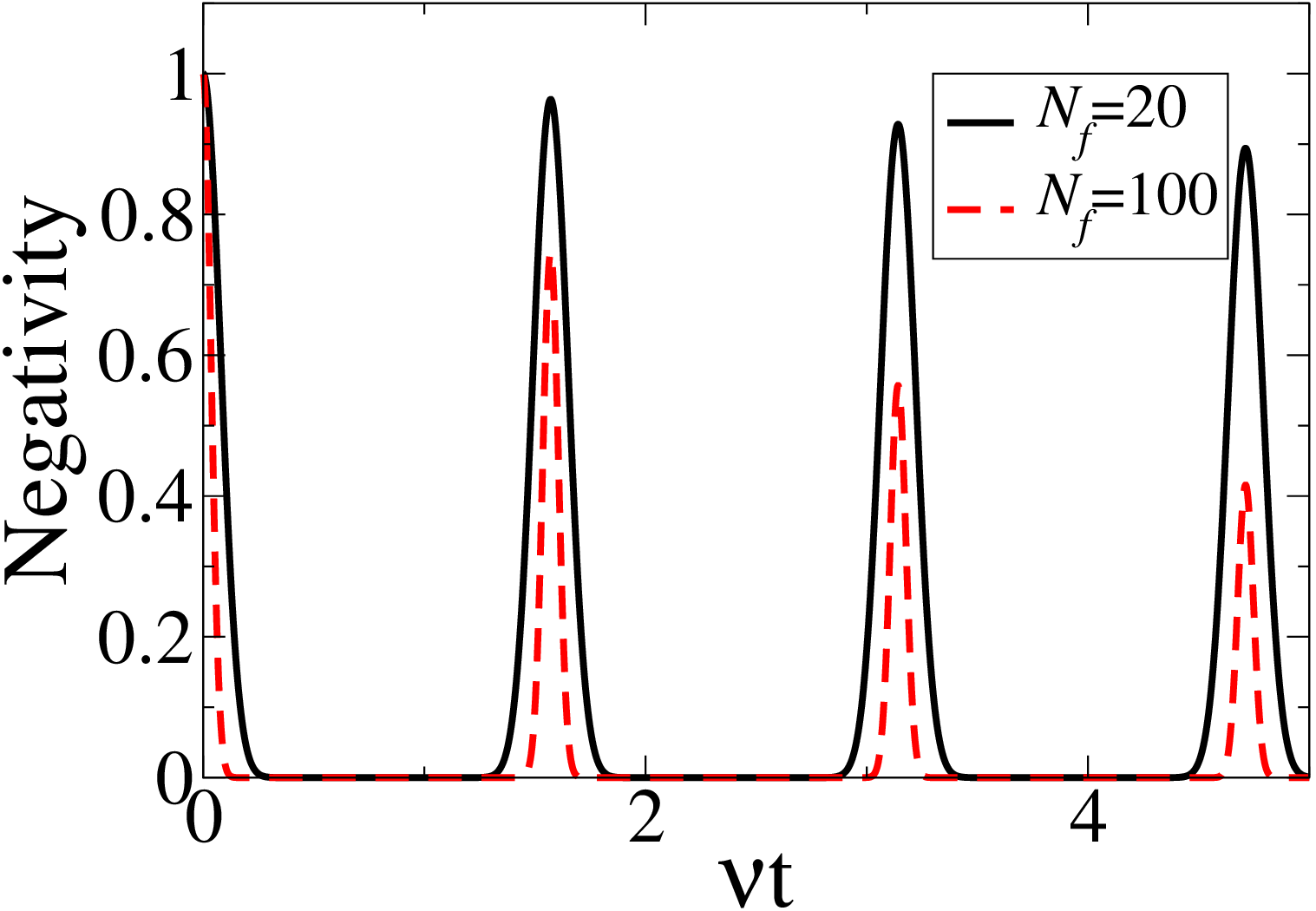}
\includegraphics*[width=0.48\columnwidth]{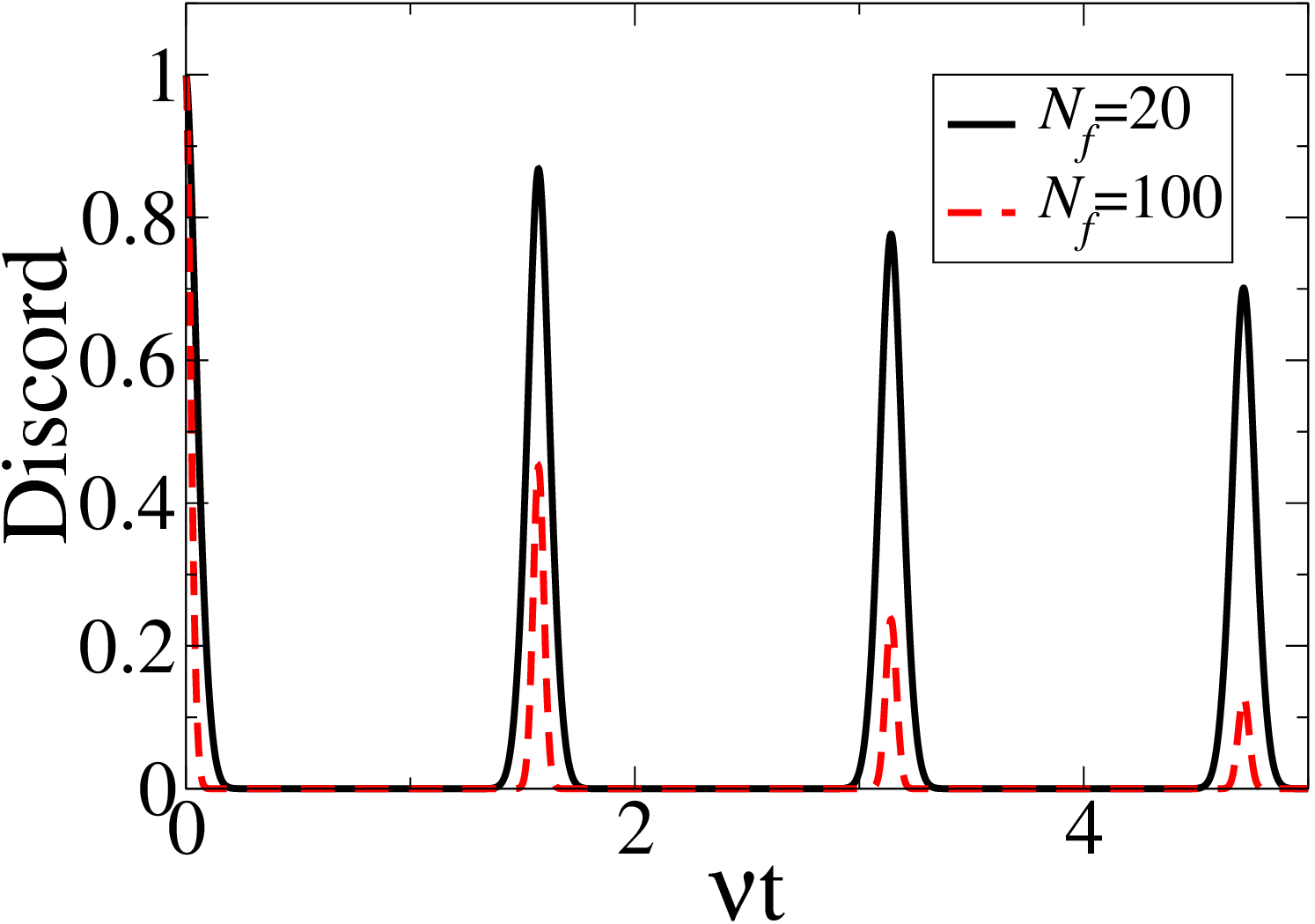}\\  
\includegraphics*[width=0.48\columnwidth]{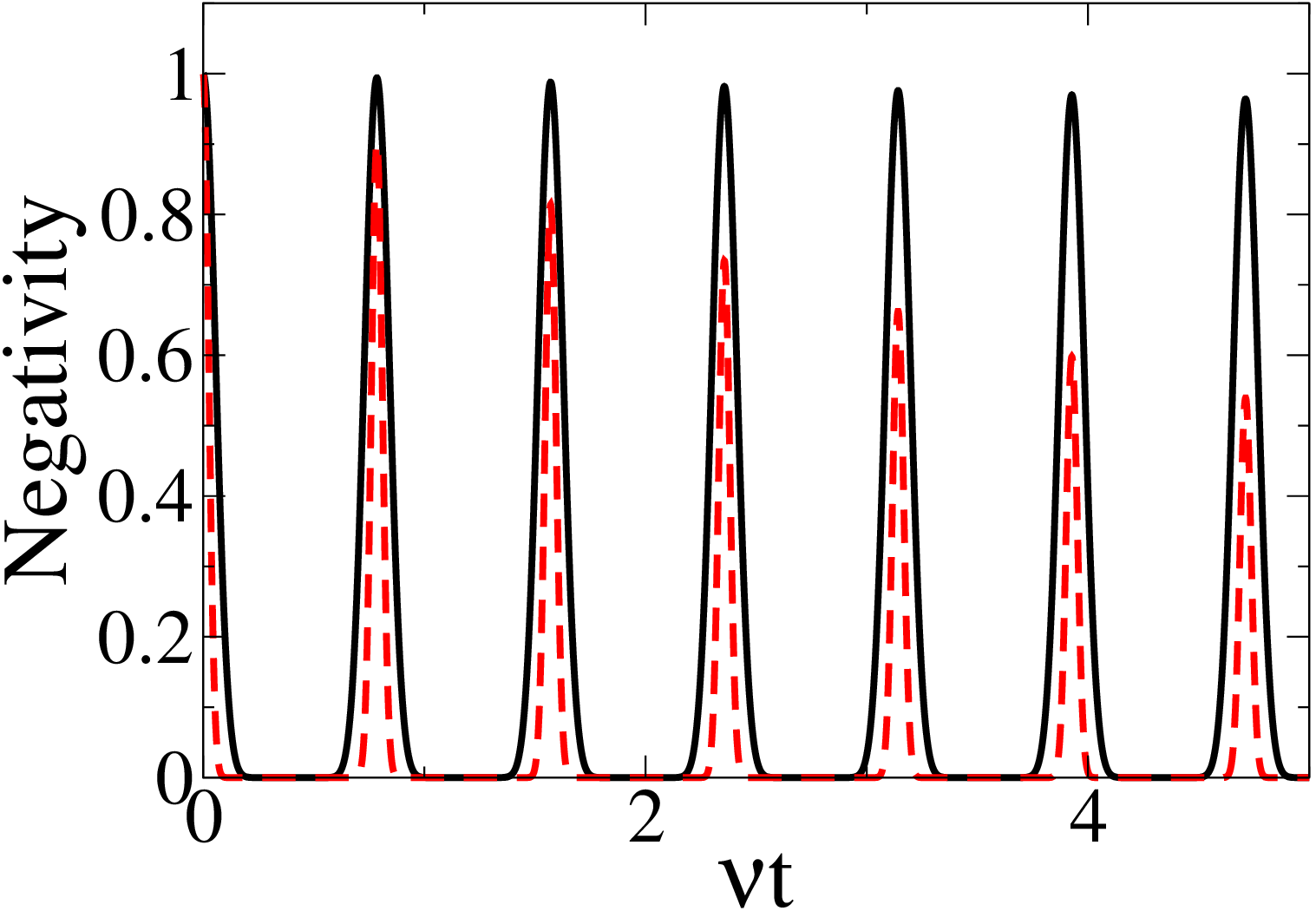}
\includegraphics*[width=0.48\columnwidth]{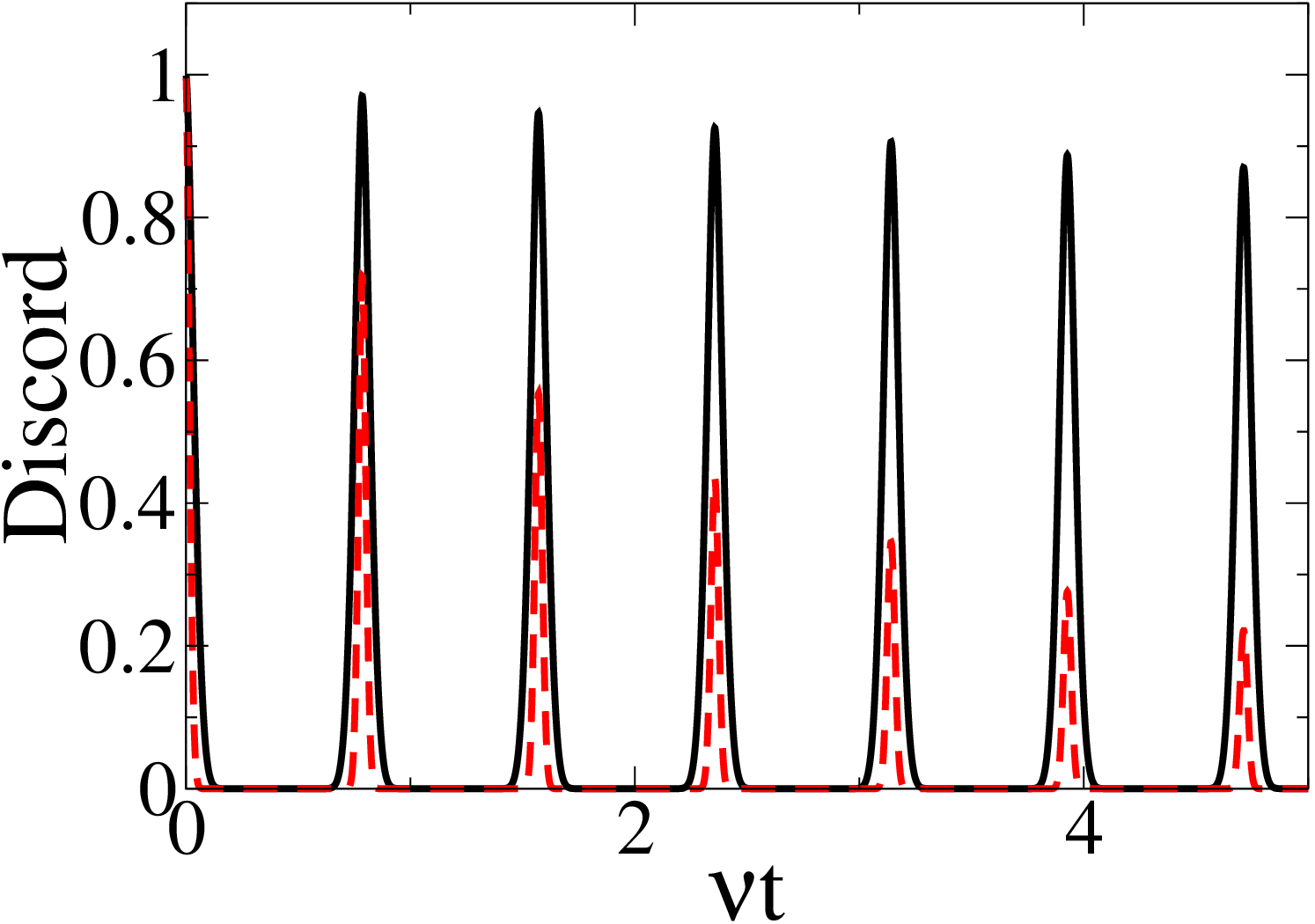}
\caption{ \label{fig4} (Color online)  
Upper panels: Time evolution  of  negativity (left) and discord (right) 
for two qubits interacting with two independent environments, consisting
in a collection of $N_f$ bistable fluctuators with spectrum $1/f^2$
when $[\gamma_1,\gamma_2]/\nu=[10^{-4},10^{4}]$. Lower 
panels: same as before, but with qubits subject to a  common
environment. } \end{figure}

In the case of two
qubits interacting with a collection of fluctuators with fixed switching
rates,  the dynamics is very different, depending on the spectrum of the
noise. The analytical expression for the negativity and the discord is
computed starting from the density matrix in Eq. \eqref{rho_many_fl}.
The negativity reads:
\begin{align}\label{negmolti}
N_{de}(t)&=|\Gamma_{de}(t)| \notag \\
N_{ce}(t)&=|\Gamma_{ce}(t)|\,,
\end{align}
where the $\Gamma$'s are given in Eq. (\ref{Gammas}).
Also in this case the density matrix preserves the X form during 
time evolution, such that the discord is calculated by 
applying the Luo formula \eqref{discordluo}:
\begin{equation} 
Q_{de(ce)}(t)= h(\Gamma_{de(ce)})\,.
\end{equation}
The negativity is the product of many oscillating coefficients, with
various periodicities. In the case of switching rates taken from a $1
/\gamma$ distribution, the product of these terms results in a monotonic
decay for both entanglement and discord. In Fig. \ref{fig3} we report
the behavior of such quantities in the case of 20 and 100 fluctuators.
As the number of fluctuators is increased the quantum correlations decay
faster. We consider 20 sources of RTN as the minimum number of
fluctuator needed to obtain both a reliable profile of the frequency
spectrum and a representative sample of the $p(\gamma)$ distribution.
Though it is possible to obtain a pink noise spectrum even with a
smaller number of fluctuators, this approximation does not describe a
sample of $1/\gamma$-distributed switching rates. 
A very different behavior arises when the $\gamma$'s are selected from a
$1/\gamma^2$ distribution, see Fig. \ref{fig4}. Phenomena of sudden death and revivals appear
for both entanglement and discord. As the number of fluctuators is
increased, the heights of the peaks decrease. The peaks have a
periodicity of $\pi/2$ and $\pi/4$ for different and common environments
respectively. As in the case of the single random fluctuator, this is
explained by considering that the selected switching rates have  small
and very close values. The product of functions with almost the same
periodicity gives a periodic behavior with  narrow peaks. 
\par
Our results
clearly show that  the  sheer knowledge of the spectrum is not
sufficient to determine the dynamical evolution of correlations. Indeed,
it is also the number of decoherence channels that plays a key role.
Different physical models of environments can lead to the same spectrum.
But, if the two-qubit system interacts with only one decoherence
channel, then revivals appear because the system is affected only by one
source of classical noise and the information can flow back. If many
sources of decoherence are present, then the information can be
completely lost, depending on the channel characteristics, that is the
distribution of the switching rates. 
\subsection{Collection of fluctuators with random switching rate}
For the sake of completeness, we
report also the expressions for negativity and discord in the case of
two qubits subject to a collection of bistable fluctuators with stochastic
switching rates. From Eq. \eqref{newcoeff}, we can write the negativity
and discord as:
\begin{align}\label{negmoltirandom}
N_{de}(t)&=\Gamma'_{de}(t)=[\Lambda_{de(ce)}]^{N_f}\\
N_{ce}(t)&=|\Gamma'_{ce}(t)|=\left|[\Lambda_{de(ce)}]^{N_f}\right|\\
Q_{de(ce)}(t)&= h(\Gamma'_{de(ce)})\,,
\end{align}
The quantum correlations decay exponentially in the case of $1/f$ noise
and with smooth damped oscillations in the case of $1/f^2$ noise. The
latter is the most general case, in which the collection of fluctuators
is described by a statistical ensemble. Also in this scenario, brown
noise leads to non-monotonic decay of negativity and discord, since all
the selected $D_{m\nu}(t)$ terms have similar periodicity and beat
effects arise, with constructive interference. 
\section{Conclusions}
We have analyzed in details the dynamics of quantum correlations, entanglement 
and discord, for a two-qubit system initially prepared in a maximally 
entangled state and then subjected to classical noise, i.e. coupled with an
external environment consisting in a collection of classical bistable 
flcutuators. In particular, due to its relevance for solid state
devices, we have addressed environments characterized by a $1/f^{\alpha}$ 
noise spectrum. We analyzed both the case of
independent environments associated to each qubit and of a common
environment, and we have taken into account explicitly the structure of
the environment.  More specifically, we have analyzed distinct
configurations of the environment, characterized by the same $1/f^{\alpha}$
spectrum, but with a different number of decoherence channels.
Analytical expressions have been found for the negativity and the
discord, showing the same qualitative behavior in all the analyzed
regimes. Indeed the choice of maximally entangled initial state 
together with the sheer dephasing nature of the interaction 
with the environment, makes discord a function of  negativity only.
In these systems, in fact, the evolved state is a mixture of Bell states.
\par
In the first configuration we have addressed, the two qubits interact with a single
bistable fluctuator, which has a random switching rate leading to an
overall $1/f^{\alpha}$ spectrum.  An oscillating behavior of quantum
correlations for both pink and brown noise has been found. The effect of a
common environment is that of better preserving the correlations and to double
the number of revivals compared to the case of independent baths. In the
presence of pink noise, the peaks of negativity and
discord decay faster than in the presence of brown noise. In fact, the
switching rates are  more evenly distributed over the range
$[\gamma_1,\gamma_2]$ than in the case of the $1/f^2$ noise, thus
leading to  massive destructive interference.  In the second
configuration the two qubits interact with a collection of bistable
fluctuators, each one with fixed switching rate chosen from a
distribution $1/\gamma^{\alpha}$; also in this case the overall noise
spectrum is $1/f^{\alpha}$.  In the case of pink noise, entanglement and
discord show a monotonic decay, while for brown noise, sudden death and
revivals occur. Quantum correlations can be written as a product of
oscillating and exponential functions. Since the switching rates of the
fluctuators are selected from a distribution $1/f^{\alpha}$, for the
pink noise they lead to destructive interference, while brown noise
enhances constructive interference.  The action of
independent or common environments has different effects on the
robustness of correlations and this agrees with previous results obtained
with different noise models \cite{Mazz,yang}.
\par
Our results
clearly show that the behavior of quantum correlations in not influenced
only by the spectrum of the environment, but also by the number of
decoherence channels, i.e. by the very structure of the environment. 
With a single decoherence channel, revivals of
correlations appear, indicating that a back-flow of information from the
environment to the system is possible. When the number of decoherence
channels is increased, the information is quickly lost, and no revivals
can occur.
\par
The characterization and the control of quantum correlations are 
fundamental for the development of quantum technology. 
Not only quantum correlations constitute a resource to
process  quantum information, but the deep understanding of their nature
will provide a better insight on the nature of quantum states themselves
and on the transition between quantum and  classical description of
physical systems.  
Our results indicate that the microscopic structure of
environment, besides its noise spectrum, is relevant for the dynamics of
quantum correlations, and may be a valid starting point for the
engineering of colored non-Gaussian environments. 
\acknowledgments
This work has been supported by MIUR (FIRB LiCHIS-RBFR10YQ3H).
MGAP thanks Sabrina Maniscalco and Elisabetta Paladino for useful
discussions. 

%

\begin{thebibliography}{10}
\bibitem{rev12}
K. Modi, A. Brodutch, H. Cable, T. Paterek, V. Vedral, 
Rev. Mod. Phys. {\bf 84}, 1655 (2012).
\bibitem{zurek}
H.~Ollivier and W.~H. Zurek,
\newblock Phys. Rev. Lett. {\bf 88}, 017901 (2001).
\bibitem{maziero}
J.~Maziero, L.~C. C\'eleri, R.~M. Serra, and V.~Vedral,
\newblock Phys. Rev. A {\bf 80}, 044102 (2009).
\bibitem{datta}
A.~Datta and S.~Gharibian,
\newblock Phys. Rev. A {\bf 79}, 042325 (2009).
\bibitem{werlang}
T.~Werlang, S.~Souza, F.~F. Fanchini, and C.~J. Villas~Boas,
\newblock Phys. Rev. A {\bf 80}, 024103 (2009).
\bibitem{datta2}
A.~Datta, A.~Shaji, and C.~M. Caves,
\newblock Phys. Rev. Lett. {\bf 100}, 050502 (2008).
\bibitem{lanyon}
B.~P. Lanyon, M.~Barbieri, M.~P. Almeida, and A.~G. White,
\newblock Phys. Rev. Lett. {\bf 101}, 200501 (2008).
\bibitem{GQD10} P.~Giorda and M.~G.~A.~Paris, 
Phys. Rev. Lett. {\bf 105}, 020503 (2010).
\bibitem{Ade10} G. Adesso and A. Datta, Phys. Rev. Lett {\bf 105},
  030501 (2010).
\bibitem{expD1} 
{R. Blandino, M. G. Genoni, J. Etesse, M. Barbieri, M. G. A. Paris, P.
Grangier, R. Tualle-Brouri}, 
Phys. Rev. Lett {\bf 109}, 180402 (2012).
\bibitem{expD2} 
M. Gu, H. M. Chrzanowski, S. M. Assad, T. Symul, K. Modi,
T. C. Ralph, V. Vedral, P. K. Lam, 
Nature Phys. {\bf 8}, 671 (2012).
\bibitem{expD3} L. S. Madsen, A. Berni, M. Lassen, U. 
L. Andersen, Phys. Rev. Lett. {\bf 109}, 030402 (2012).
\bibitem{giulini}
D.~Giulini and \emph{et~al.},
\newblock {\em Decoherence and the Appearance of a Classical World in Quantum
  Theory} (Springer, 1996).
\bibitem{Zurek2}
W.~H. Zurek,
\newblock Rev. Mod. Phys. {\bf 75}, 715 (2003).
\bibitem{hu1}
M.-L. Hu and H. Fan,
Ann. Phys. {\bf 327} 851 (2012);
M.-L. Hu, H. Fan
Ann. Phys. 327(9): 2343 - 2353 (2012)

\bibitem{cvqc1}
S. Maniscalco, S. Olivares, M. G. A. Paris, 
Phys. Rev. A {\bf 75}, 062119 (2007); 
R. Vasile, S. Olivares, M. G. A. Paris, S. Maniscalco, 
Phys. Rev. A {\bf 80}, 062324 (2009).
\bibitem{cvqc2}
R. Vasile, P. Giorda, S. Olivares, M. G. A. Paris, S.Maniscalco, 
Phys. Rev. A {\bf 82}, 012313 (2010).
\bibitem{vasile}
{R. Vasile, S. Olivares, M. G. A. Paris, S. Maniscalco}
 Phys. Rev. A {\bf 83}, 042321 (2011).
\bibitem{cvqc3}
M. G. Genoni, P. Giorda, M. G. A. Paris, 
Phys. Rev. A {\bf 78}, 032303 (2008); 
G. Brida, I. P. Degiovanni, A. Florio, M. Genovese, P.
Giorda, A. Meda, M. G. A. Paris, A. Shurupov, 
Phys. Rev. Lett. {\bf 104}, 100501 (2010).
\bibitem{cvqc4} A. Ferraro, M. G. A. Paris,
Phys. Rev. Lett {\bf 108}, 260403 (2012).
\bibitem{Warren}
W.~S. Warren,
\newblock Science {\bf 277}, 1688 (1997).
\bibitem{Oliveira}
J. G. Filgueiras, T. O. Maciel, R. E. Auccaise, R. O. Vianna, R. S.
Sarthour, I. S. Oliveira, Q. Inf. Proc. {\bf 11}, 1883 (2012).

\bibitem{buscemitelep}
F.~Buscemi, P.~Bordone, and A.~Bertoni,
Phys. Rev. B {\bf 81}, 045312 (2010);
F.~Buscemi,
 Phys. Rev. A {\bf 83}, 012302 (2011);
F.~Buscemi, P.~Bordone and A.~Bertoni,J. Phys.: Condens. Matter {\bf 21} 305303 (2009);
F.~Buscemi, P.~Bordone and A.~Bertoni, Eur. Phys. J. D 66, 312 (2012).

\bibitem{buscemidem}
F.~Buscemi, P.~Bordone, and A.~Bertoni,
 New J. Phys. {\bf 13}, 013023 (2011);
F.~Buscemi, P.~Bordone, and A.~Bertoni,
 Phys. Rev. B {\bf 76}, 195317 (2007).

\bibitem{Sarovar}
M.~Sarovar, A.~Ishizaki, G.~R. Fleming, and K.~B. Whaley,
\newblock Nature Phys. {\bf 6}, 462 (2010).
\bibitem{Fassioli}
F.~Fassioli and A.~Olaya-Castro,
\newblock New J. Phys. {\bf 12}, 085006 (2010).

\bibitem{yu}
T. Yu and J.H. Eberly, {\it Science} {\bf 323} (2009) 598.
\bibitem{yu2}
T. Yu and J.H. Eberly, {\it Phys. Rev. Lett.} {\bf 93} (2004) 140404.
\bibitem{yang}
Q.~Yang, M.~Yang, D.~C. Li, and Z.~L. Cao,
\newblock Int. J. Theor. Phys. {\bf 51}, 2160 (2012).

\bibitem{cao}
Z.-L. Cao and H.~Zheng,
\newblock Eur. Phys. J. B {\bf 68}, 209 (2009).

\bibitem{s1}
L. Mazzola, J. Piilo and S. Maniscalco, Phys. Rev. Lett. {\bf 104},
200401 (2010).

\bibitem{ma}
J.~Ma, Z.~Sun, X.~Wang, and F.~Nori,
\newblock Phys. Rev. A {\bf 85}, 062323 (2012).

\bibitem{Ficek}
Z.~Ficek and R.~Tana\ifmmode~\acute{s}\else \'{s}\fi{},
\newblock Phys. Rev. A {\bf 74}, 024304 (2006).

\bibitem{Mazz}
L.~Mazzola, S.~Maniscalco, J.~Piilo, K.-A. Suominen, and B.~M. Garraway,
\newblock Phys. Rev. A {\bf 79}, 042302 (2009).

\bibitem{Comp2}
B.~Bellomo, R.~Lo~Franco, and G.~Compagno,
\newblock Phys. Rev. Lett. {\bf 99}, 160502 (2007).

\bibitem{zhang12}
Y.-J. Zhang, W. Han, C.-J. Shan, and Y.J. Xia
\newblock J. Opt. Soc. Am. B{\bf 28}, 2060 (2012)

\bibitem{lofranco}
R.~Lo~Franco, B.~Bellomo, E.~Andersson, and G.~Compagno,
\newblock Phys. Rev. A {\bf 85}, 032318 (2012).

\bibitem{Zhou}
D.~Zhou, A.~Lang, and R.~Joynt,
\newblock Quantum Inf. Process. {\bf 9}, 727 (2010).

\bibitem{pal12}
A. D'Arrigo, R. Lo Franco, G. Benenti, E. Paladino, and G. Falci
\newblock arXiv:1210.1122.

\bibitem{s2} L. Mazzola, J. Piilo and S. Maniscalco, Int. J. Quant. Inf.
{\bf 9}, 981 (2011).

\bibitem{wold}
H.~L. Wold, H.~Brox, Y.~M. Galperin, and J.~Bergli,
\newblock arXiv:1206.2174v1 .

\bibitem{Z2}
A.~De, A.~Lang, D.~Zhou, and R.~Joynt,
\newblock Phys. Rev. A {\bf 83}, 042331 (2011).

\bibitem{benedetti2}
P.~Bordone, F.~Buscemi, and C.~Benedetti,
\newblock Fluct. Noise Lett. {\bf 11}, 1242003 (2012).

\bibitem{Lof2}
R.~L. Franco, A.~D'Arrigo, G.~Falci, G.~Compagno, and E.~Paladino,
\newblock Phys. Scripta {\bf T147}, 014019 (2012).

\bibitem{Duty}
T.~Duty, D.~Gunnarsson, K.~Bladh, and P.~Delsing,
\newblock Phys. Rev. B {\bf 69}, 140503 (2004).

\bibitem{eroms}
J. Eroms, L. C. van Schaarenburg, E. F. C. Driessen, J. H. Plantenberg,
C. M. Huizinga, R. N. Schouten, A. H. Verbruggen, C. J. P. M. Harmans, and J. E. Mooij, 
\newblock App. Phys. Lett. {\bf 89}, 122516 (2006).

\bibitem{parman91}
C. E. Parman, N. E. Israeloff and J. Kakalios
 \newblock Phys. Rev. B {\bf 44}, 8391 (1991).

\bibitem{rogers84}
C. T. Rogers and R. A. Buhrman
 \newblock Phys. Rev. Lett.  {\bf 53}, 1272 (1984).

\bibitem{rogers85}
C. T. Rogers and R. A. Buhrman
\newblock Phys. Rev. Lett.  {\bf 55}, 859 (1985).

\bibitem{peters99}
M. Peters, J. Dijkhuis and L. Molenkamp
\newblock J. Appl. Phys.  {\bf 86}, 1523 (1999).

\bibitem{weissman}
M. B. Weissman,
\newblock Rev. Mod. Phys. {\bf 60}, 537 (1998).

\bibitem{meno}
K. Kakuyanagi, T. Meno, S. Saito, H. Nakano, K. Semba, H.
Takayanagi, F. Deppe, 
A. Shnirman, \newblock Phys. Rev. Lett. {\bf 98}, 047004 (2007).

\bibitem{tsai}
F.~Yoshihara, K.~Harrabi, A.~O. Niskanen, Y.~Nakamura, and J.~S. Tsai,
\newblock Phys. Rev. Lett. {\bf 97}, 167001 (2006).

\bibitem{paladino}
E.~Paladino, L.~Faoro, G.~Falci, and R.~Fazio,
 Phys. Rev. Lett. {\bf 88}, 228304 (2002);
G.~Falci, A.~D'Arrigo, A.~Mastellone, and E.~Paladino,
 Phys. Rev. Lett. {\bf 94}, 167002 (2005).


\bibitem{Fac2}
E.~Paladino, A.~D'Arrigo, A.~Mastellone and G.~Falci
New J. Phys. {\bf 13} 093037 (2011);
B.~Bellomo, G.~Compagno, A.~D'Arrigo, G.~Falci, R.~Lo Franco, and E.~Paladino,
Phys. Rev. A {\bf 81}, 062309 (2010) .

\bibitem{Bukard}
G.~Burkard,
\newblock Phys. Rev. B {\bf 79}, 125317 (2009).

\bibitem{nota}
The term colored is a metaphor to suggest that white noise,
in analogy with white light, comes from an environment which contains all different
frequencies with a flat power spectrum, whereas other colors of noise can
be obtained by selecting specific frequency ranges. 

\bibitem{galperin06}
Y. M. Galperin, B. L. Altshuler, J. Bergli, D. V. Shantsev, 
\newblock Phys. Rev. Lett. {\bf 96}, 097009, (2006).

\bibitem{moe07}
D. L. Moehring, P. Maunz, S. Olmschenk, K. C. Younge1, 
D. N. Matsukevich, L.-M. Duan, and C. Monroe,
\newblock Nature  {\bf 449}, 68 (2007).

\bibitem{sillanpa07}
M. A. Sillanpaa, J. I. Park and R. W. Simmonds, 
\newblock Nature  {\bf 449}, 438 (2007).

\bibitem{majer07}
J. Majer, J. M. Chow, J. M. Gambetta, J. Koch, 
B. R. Johnson, J. A. Schreier, L. Frunzio, D. I. Schuster, A. A. Houck, 
A. Wallraff, A. Blais, M. H. Devoret, S. M. Girvin and R. J. Schoelkopf, 
\newblock Nature {\bf 449}, 443 (2007).

\bibitem{ludvi84}
A. Ludviksson, R. Kree, and A. Schmid,
\newblock Phys. Rev. Lett., {\bf 52}, 950 (1984).
\bibitem{kogan84}
S. M. Kogan and K. E. Nagaev,
\newblock Solid State Comm., {\bf 49}, 387 (1984).
\bibitem{mx1}
D. A. Averin, Sol. State Comm. {\bf 105}, 659 (1998).
\bibitem{mx2}
L. B. Ioffe, V. B. Geshkenbein, M. V. Feigel'man, A. L. Fauch\'ere, 
G. Blatter, Nature {\bf 398}, 679 (1999).
\bibitem{mx3} J.E. Mooij, T. P. Orlando, 
L. Levitov, L. Tian, C. H. van der Wal, S, Lloyd, 
Science {\bf 285}, 1036 (1999).
\bibitem{mx4}
G. Falci, R. Fazio, G. M. Palma, J. Siewert1, V. Vedral, 
Nature {\bf 407}, 355 (2000).

\bibitem{Joh00} R. E. Johanson, M. Gunes, S.O. Kasap, 
J. Non-Cryst. Sol. {\bf 266-269}, 242 (2000).
\bibitem{Raq00} B. Raquet, A. Anane, S. Wirth, P. Xiong, and S. von
Moln\`ar, Phys. Rev. Lett. {\bf 84}, 4485-4488 (2000).
\bibitem{Jun03} 
 G. Jung, Y. Paltiel,E. Zeldov, Y. Myasoedov, M. L. Rappaport, 
 M. Ocio, S. Bhattacharya, M. J. Higgins
 Proc. SPIE {\bf 5112}, 222 (2003).
\bibitem{negat}
G.~Vidal and R.~F. Werner,
\newblock Phys. Rev. A {\bf 65}, 032314 (2002).

\bibitem{neg_trip}
F.~Buscemi and P.~Bordone,
\newblock Phys. Rev. A {\bf 84}, 022303 (2011).

\bibitem{vedral}
L.~Henderson and V.~Vedral,
\newblock J. Phys. A {\bf 34}, 6899 (2001).

\bibitem{luo}
S.~Luo,
\newblock Phys. Rev. A {\bf 77}, 042303 (2008).

\bibitem{bergli}
J.~Bergli, Y.~M. Galperin, and B.~L. Altshuler,
\newblock New J. Phys. {\bf 11}, 025002 (2009).

\bibitem{benedetti3}
C.~Benedetti, F.~Buscemi, P.~Bordone, and M.~G.~A. Paris,
\newblock Int. J. Quant. Inf. {\bf 10}, 1241005 (2012)

\bibitem{benedetti1}
C.~Benedetti, F.~Buscemi, and P.~Bordone,
\newblock Phys. Rev. A {\bf 85}, 042314 (2012).


\bibitem{daf04}
S. Daffer, K. Wodkiewicz, J. D. Cresser and J. K. McIver, Phys. Rev. A
{\bf 70}, 010304 (2004).

\bibitem{kuopo}
P. Kuopanportti, M. 
M\"{o}tt\"{o}nen, V. Bergholm, O-P. Saira, J. Zhang, K. BirgittaWhaley, 
\newblock Phys. Rev. A {\bf 77}, 032334 (2008).

\end{thebibliography}
\end{document}